\documentclass[preprint]{aastex}

\newcommand{\bdv}[1]{\mbox{\boldmath$#1$}}
\def\bpi{{\bdv\pi}}
\def\bmu{{\bdv\mu}}
\def\E{{\rm E}}
\def\L{{\rm L}}
\def\S{{\rm S}}
\def\rel{{\rm rel}}

\def\Spitzer{{\em Spitzer}}
\def\Kepler{{\em Kepler}}
\def\K2{{\em K2}}
\def\WFIRST{{\em WFIRST}}

\title{Lens Masses and Distances from Microlens Parallax and Flux}

\author{Jennifer C. Yee}
\affil{Sagan Fellow; Harvard-Smithsonian Center for Astrophysics, 60 Garden St., Cambridge, MA 02138, USA}

\begin{document}

\begin{abstract}
I present a novel method for measuring lens masses for microlensing
events. By combining a measured lens flux with the microlens parallax
$\vec{\bpi}_\E$, it is possible to derive the mass of the lens system
without knowing the angular size of the Einstein ring,
$\theta_\E$. This enables mass and distance measurements for single,
luminous lenses, as well as binary and planetary lenses without
caustic crossings. I discuss applications of this method in the
contexts of the \Spitzer, \Kepler, and \WFIRST\, microlensing
missions.
\end{abstract}


{\section{Introduction}
\label{sec:intro}}

Microlensing is a powerful technique because it can probe a wide
variety of systems including isolated black holes
\citep[e.g. ][]{Poindexter05} and planetary systems distant from the
Earth \citep[e.g. ][]{Udalski15}. It probes lens systems regardless of
their light because it uses light from a background source to create
the signal. The flip side of this is that it is very difficult to
determine the absolute properties of the lens system. For example,
MOA-2011-BLG-262 has a 2-body lens with a planetary mass ratio
$q\sim5\times10^{-4}$, but the solution is ambiguous between an
M-dwarf with a gas giant planet and a free-floating planet with a well
separated sub-Earth mass moon \citep{Bennett14}.

The reason it is so difficult to determine the properties of the lens is that its mass $M_\L$, distance $D_\L$, and motion relative to the source $\mu_{\rel}$ are all encoded in the primary microlens observable 
\begin{equation}
 t_\E = \frac{\theta_\E}{\mu_{\rel}}, 
\label{eqn:te}
\end{equation}
where 
\begin{equation}
\theta_\E = \sqrt{\kappa M_\L\pi_{\rel}} \quad {\rm where} \quad \pi_{\rel} = {\rm AU\,}\left(\frac{D_\L-D_\S}{D_\S D_\L}\right) \quad {\rm and} \quad \kappa = 8.14\, \frac{\rm mas}{M_{\odot}}.
\label{eqn:thetae}
\end{equation}
$D_\S$ is the distance to the source, which is assumed to be in the
Bulge (i.e. at $\sim 8$ kpc). Hence, two additional measurements are
needed to determine the lens mass and distance.

Historically, measurements of the lens mass have focused on combining
a measurement for the angular size of the Einstein ring, $\theta_\E$,
with other information. $\theta_\E$ is often measured for 2-body
microlensing events. This is because 2-body microlenses are identified
as such by interactions of the source light with the lens caustic
structure. The caustic is a closed curve at which the magnification
formally diverges to infinity. As a result, if the source crosses a
caustic, the observed magnification depends sensitively on the
physical size of the source, which may be expressed as $\rho =
\theta_{\star}/\theta_\E$ where $\theta_{\star}$ is the angular size
of the source. Hence, if the lens is observed to have two bodies, it
is likely that a caustic crossing has been observed and thus, that $\rho$
is measured, which leads to a measurement of $\theta_\E$ once
$\theta_{\star}$ is determined from the source color and magnitude
\citep{Yoo04}.

Once $\theta_{\E}$ is measured, this yields a mass-distance
relationship for the lens star such as those shown by the blue lines
in Figures \ref{fig:ob0124} and \ref{fig:wfirst}. If this can be
combined with a second mass-distance relationship, then the mass and
distance to the lens are determined. This may be done in one of two
ways. First, if parallax effects are measured in the light curve, the
measurement of the microlens parallax vector $\vec{\bpi}_\E$ directly gives the lens mass and distance:
\begin{equation}
M_\L = \frac{\theta_\E}{\kappa\pi_\E} ;\quad D_\L = \left(\pi_\E\theta_\E - \pi_\S\right)^{-1},
\label{eqn:mass}
\end{equation}
where $\pi_\S=D_\S^{-1}$. The second method is to make a measurement
of the lens flux using high resolution imaging so that stars unrelated
to the microlensing event are resolved. This may be done either while
the lens and source are still superposed or after they have separated,
depending on the brightness of the source. This flux measurement then
gives a magnitude-distance relationship that can be compared to the
mass-distance relationship from $\theta_\E$ via stellar isochrones,
e.g. the magenta lines in Figure \ref{fig:ob0124} and
\ref{fig:wfirst}.

However, there are many cases in which $\theta_\E$ is not
measured. This includes almost all point lenses, for which a caustic
crossing requires that the lens transit the face of the source, and
also 2-body systems without caustic crossings, a situation that is not
uncommon for stellar binary lenses. For these systems, I propose an
alternate means of determining the lens mass (and distance) from the
parallax.

\section{Masses from Parallax and Flux}

In detail, the microlens parallax, $\vec{\bpi}_\E$, is measured from
the apparent displacement as a function of time of the lens relative
to the source due to the parallax effect as compared to what would be
observed from rectilinear motion as seen from a single location. For
example, two observers at different locations will see a different
alignment for the source and lens, and therefore a different projected
separation. In addition, as with all separations in microlensing, this
displacement is measure relative to the size of the Einstein
ring. Hence, the microlens parallax is a vector that depends on the
lens-source relative parallax, $\pi_{\rel}$, the direction of the
lens-source relative proper motion, $\vec{\bmu}_{\rel}/\mu_{\rel}$,
and is scaled to the size of the Einstein ring, $\theta_\E$:
\begin{equation}
\vec{\bpi}_\E = \frac{\pi_{\rel}}{\theta_\E}\frac{\vec{\bmu}_{\rel}}{\mu_{\rel}}.
\label{eqn:pie}
\end{equation}

The basic idea of measuring masses from parallax and flux stems from
Equation \ref{eqn:mass}. If a measurement of the lens flux plus a
measurement of $\theta_{\E}$ can lead to a measurement of the lens
mass, so too must a measurement of the lens flux plus a measurement of
$\vec{\bpi}_\E$ lead to a measurement of the lens mass. This is just
the third possible pairing of the luminosity-distance relationships
shown in Figures \ref{fig:ob0124} and \ref{fig:wfirst}. 

Previously, \citet{Dong09_071} combined constraints from several
microlensing effects, the strongest of which was parallax, with an
{\em HST} measurement of the lens flux to determine the mass of
OGLE-2005-BLG-071. However, there has not been a measurement of a lens
mass from only parallax and lens flux.

The reasons this idea has never before been applied in this form are
two-fold. First, most of the microlensing events that have been
published and for which absolute (rather than statistical) masses are
necessary have been 2-body lenses, especially star-planet systems. In
those cases, $\theta_\E$ is almost always measured, so the techniques
described in Section \ref{sec:intro} are sufficient. Second, before
the advent of space-based microlensing, microlens parallaxes were
rarely measured, so the most frequent situation was one with a
measurement of $\theta_\E$ but no measurement of $\vec{\bpi}_\E$
\citep[the first space-based parallax measurement was made by
][]{Dong07}.

However, the \Spitzer\, microlens parallax campaign (and soon the \K2
microlensing campaign) has enabled highly precise measurements of the
microlens parallax for a significant number of events. In fact, in the
case of planetary microlens OGLE-2014-BLG-0124L, the limiting factor
in determining the lens mass was the precision of $\theta_\E$ rather
than any uncertainty in the microlens parallax. In addition, \WFIRST\,
will measure at least one component of the parallax vector extremely
precisely for large numbers of events as well as measuring lens
fluxes, so the combination of those two pieces of information will
yield more detail about the nature of the lens. In the next section,
I give additional details about how this idea can be applied to these
space missions.

\section{Specific Applications}

\subsection{Satellite Parallaxes}

\begin{figure}
\plotone{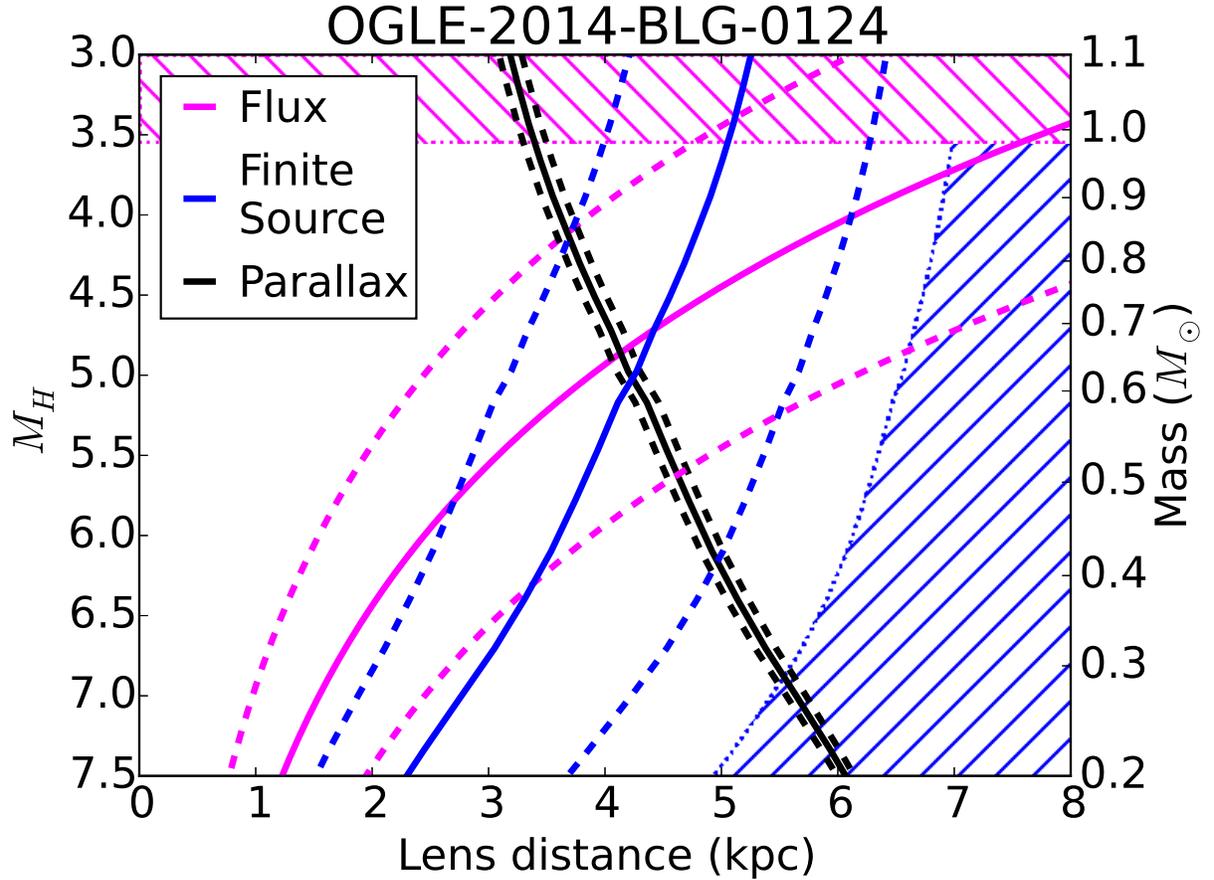}
\caption{Absolute Magnitude and Mass constraints on OGLE-2014-BLG-0124L, a
  planetary system from the 2014 \Spitzer\, campaign. For this
  microlensing event, the parallax measured from the difference in the
  light curves as seen from \Spitzer\, and the Earth is extremely
  precise, yielding the mass-distance relation shown in
  black. However, the source size was not well measured from finite
  source effects, leading to a large uncertainty in the mass of the
  system. The light curve itself yields only an upper limit on the
  source size, which excludes the blue hatched region, while upper
  limits on the lens flux exclude the magenta hatched region. The blue
  curves show the mass-distance relation from the nominal value of
  $\rho$ from \citet{Udalski15}, but anything in the non-excluded
  region is consistent with the light curve at the 3-$\sigma$
  level. The magenta line shows the hypothetical constraints from a
  measurement of the lens flux assuming that the lens star is an $0.71
  M_{\odot}$ star located at 4.1 kpc as in the \citet{Udalski15}
  solution. 1-$\sigma$ uncertainties are shown as the dashed lines.}
\label{fig:ob0124}
\end{figure}

Microlens parallax measurements from simultaneous observations from
the Earth and a satellite in solar orbit (such as \Spitzer\, or
\Kepler), can yield very precise measurements of the parallax
\citep{CalchiNovati15,Udalski15,Yee15a,Zhu15}, albeit subject to the
four-fold degeneracy \citep{Refsdal66,Gould94}. If this degeneracy can
be resolved, the resulting mass-distance relationship is well-defined
(e.g. black lines in Figure \ref{fig:ob0124}.

The four-fold degeneracy may be resolved in several ways. First, note
that the four-fold degeneracy arises from a two-fold degeneracy in
magnitude and a two-fold degeneracy in direction. Of these, only the
degeneracy in magnitude affects the measured lens mass and
distance. For high-magnification microlensing events, the difference
in the magnitude of $\vec{\bpi}_\E$ for the different solutions is small
enough that it may be ignored \citep{GouldYee12}. In addition,
higher-order effects in the light curve can break the degeneracy
outright. Alternatively, \citet{CalchiNovati15} showed that the
degeneracy may be broken in some cases using the Rich argument. This
is a statistical argument that if the two components of the parallax
vector are highly unequal in one case and close to equal in the other,
the latter is more likely \citep[see ][for a full
  explanation]{CalchiNovati15}. Finally, other ideas exist for
breaking this degeneracy in specific circumstances
\citep[e.g. ][]{Gould13,Yee13,Yee15d}.

Figure \ref{fig:ob0124} provides a specific example from the 2014
\Spitzer\, microlens parallax campaign. \citet{Udalski15} measured the
parallax for this event, but finite source effects were not
observed. However, they were able to place an upper limit on $\rho$
based on the fact that a larger source would have led to observable
effects on the light curve, and they were able to set a lower limit on
$\rho$ from an upper limit on the lens flux. Based on these
constraints, they estimated the mass and distance of this
planetary system.

Using the 4.0 Gyr, solar metallicity isochrone from \citet{An07} and
assuming the source is at 8 kpc, I transform the two mass-distance
relations from parallax and finite source effects into absolute
magnitude-distance relations as shown in the figure. Note that the
relation based on finite source effects is derived from the nominal
value and uncertainty in $\rho$ quoted in Table 1 of
\citet{Udalski15}. However, their analysis finds that $\rho$ is more
or less uniformly distributed between the constraints described
above. I also construct the absolute magnitude-distance relation that
would result from a direct measurement of the lens flux, assuming a
$0.71 M_{\odot}$ lens at $4.1$ kpc and that the extinction varies
linearly with distance and has a total value $A_H=0.399$
\citep{SchlaflyFnkbeiner11}. The uncertainty in this relation assumes
that the source and lens are still superposed, so that the lens flux
must be derived from a measurement of their combined flux, and that
the uncertainty in this flux measurement in 0.05 mag.  Note that this
figure shows just one of the two solutions from \citet{Udalski15}, but
the other solution is almost identical.

Because of the uncertainty in $\rho$, a flux measurement of
OGLE-2014-BLG-0124L, combined with the known parallax, would yield a
direct measurement of the lens (and planet) mass in this system.

In addition, most microlenses from the \Spitzer\, sample have 
parallax measurements but no measurement of finite source effects
because they are single objects. For these objects,
\citet{CalchiNovati15} were only able to make statistical estimates of
the lens distances from the measured parallaxes through a kinematic
argument that assumes $\mu_{\rel}$ is approximately known based
on the idea that the Sun and the lens are moving together. Hence, a
flux measurement is the only way to provide direct measures of the
lens masses and distances. One specific example is OGLE-2015-BLG-1285,
for which a flux measurement of the lens secondary would lead to a
mass measurement for the system and therefore determine whether the
primary is a black hole or a neutron star
\citep{Shvartzvald15}. Moreover, direct measurements of the lens
fluxes would improve the measurements of their distances, which would
in turn improve the measurement of the Galactic distribution of planets.

\subsection{WFIRST}

This method of combining lens flux and parallax measurements is also
relevant for measuring lens masses and distances with
\WFIRST. However, in this case, there is more information
available. \WFIRST\, will make three measurements that constrain the
lens mass. First, because of its higher resolution, the microlensing
observations will resolve out blended background stars. Hence, any
light left over will be due to the lens, a companion to the lens, or a
companion to the source. In general, the relative probabilities of
these various scenarios can be calculated, so lens flux measurements
will be routine. Second, the precision of \WFIRST\, will allow the
measurement of astrometric microlensing effects, which gives a
measurement of $\theta_{\E}$ \citep{GouldYee14}. Finally, \WFIRST\,
will measure parallaxes from the orbital motion of the satellite about
the Sun. Because the events are short, orbital parallax measurements
are primarily sensitive to the parallel component of the microlens
parallax vector $\pi_{\E, \parallel} = \vec{\bpi}_{\E}\sin\lambda$,
where $\lambda$ is the latitude of the event with respect to the
ecliptic \citep{Gould13}. This leads to a 1-D measurement of the
parallax \citep{GouldMB94}, a problem which is exacerbated by the fact
that the ecliptic runs through the Galactic Bulge. However, if the
parallax is measured better, e.g.  because the parallax is large or
more complex parallax effects are observed
\citep[cf.][]{Gould13,Yee13}, this measurement of the parallax is
quite powerful because it takes a completely different form from the
other relations.

\begin{figure}
\plottwo{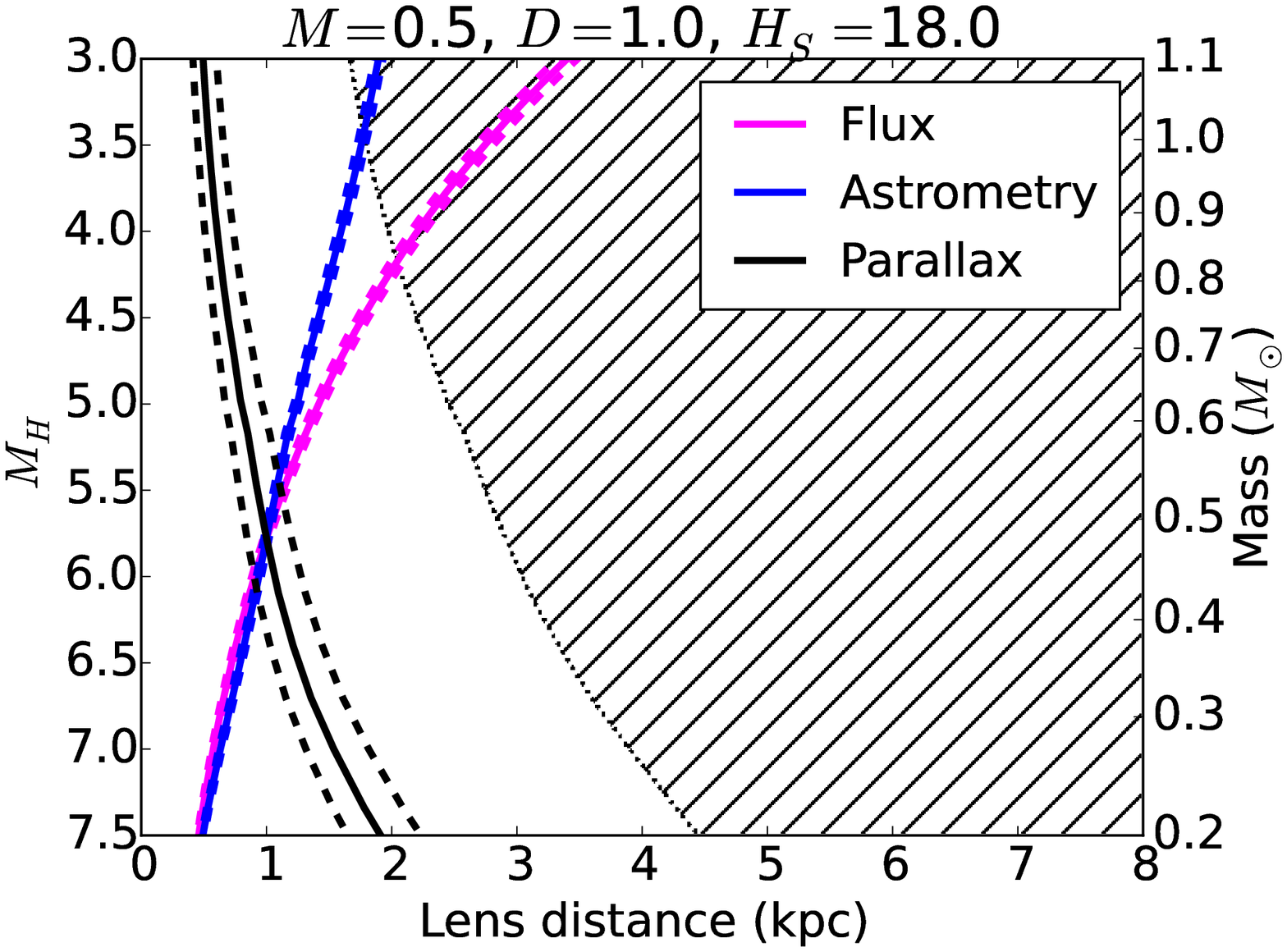}{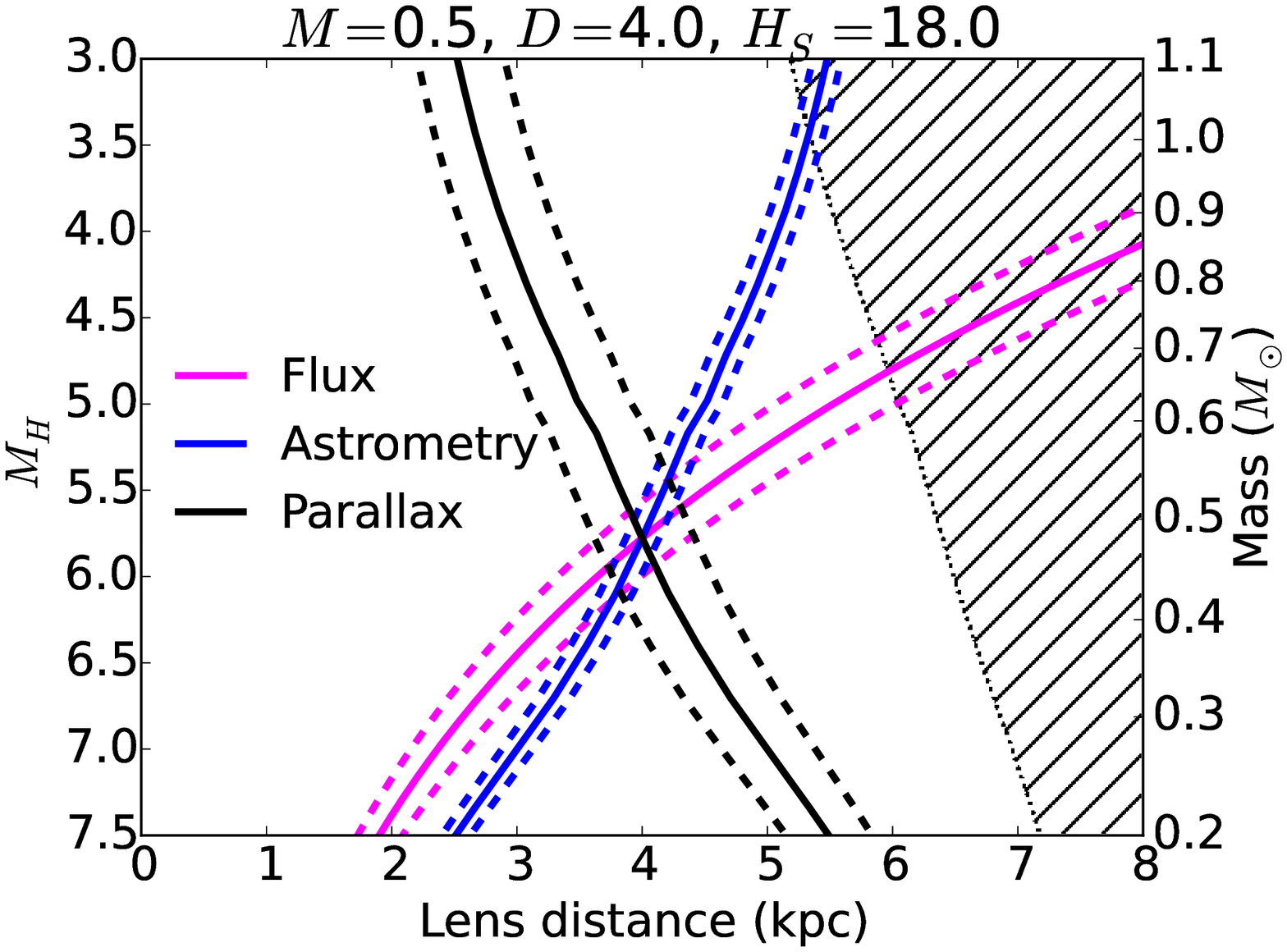}
\caption{Absolute Magnitude-Distance relations for a $0.5 M_{\odot}$
  star at 1.0 kpc (left) and 4.0 kpc (right). \WFIRST\, will measure
  three different constraints: the flux of the lens (magenta),
  astrometric microlensing (blue), and parallax (black). For
  parallaxes, \WFIRST\, will be much more sensitive to the parallel
  component $\pi_{\E,\parallel}$ than to the perpendicular component
  $\pi_{\E,\perp}$. The hatched region shows the region that is ruled
  out if only $\pi_{\E,\parallel}$ is measured (assuming
  $\arctan{\pi_{\E,\parallel}/\pi_{\E,\perp}}=30^{\circ}$); the dashed
  lines show the 1-$\sigma$ uncertainties if the parallax is measured
  to 10\%.}
\label{fig:wfirst}
\end{figure}

Figure \ref{fig:wfirst} illustrates the interplay between these three
measurements for a typical case of a $0.5 M_{\odot}$ lens star at 4.0
kpc and a case in which the lens is much closer ($D_{\rm L} = 1.0$
kpc). I have assumed the source is a dwarf star at 8 kpc with $H=18.0$
mag, known with a precision of $0.05$ mag from the microlensing
model. For the purposes of measuring the flux of the lens, I have
adopted an uncertainty in the calibrated flux at baseline of $0.05$
mag and assumed linearly varying extinction with a total value of
$A_H=0.4$. For the measurement of $\theta_{\E}$ from microlens
astrometry, I have used Equation 18 from \citet{GouldYee14} and
adopted their fiducial parameters (i.e., $\sigma_{\rm phot} = 0.01$,
FWHM$=175$ mas, $N=7000$, and $\beta=0.7$). Finally, for the parallax,
I show two cases. The hatched regions show the region excluded if only
1-D parallaxes are measured (with $\lambda=30^\circ$), while the dashed
lines assume a 10\% uncertainty in the total magnitude of the parallax
vector.

This figure clearly shows that what can be learned from the
combination of lens flux and parallax depends on how well the parallax
is measured and somewhat on the orientation of $\vec{\bpi}_\E$ (as $\lambda
\rightarrow 90^\circ$, $\pi_{\E,\parallel} \rightarrow \pi_{\E}$, so
the constraints from 1-D parallaxes improve). However, the subset of
cases for which the parallax is measured are important for validating
the \WFIRST\, results.

The most striking thing about Figure \ref{fig:wfirst} is that the
luminosity-distance relationship derived from the parallax has a
completely different form from that derived from $\theta_\E$. This
provides an important consistency check for the derived mass,
especially as the relations from $\theta_\E$ and flux become
degenerate for low-mass, nearby lenses (see left panel of Figure
\ref{fig:wfirst}). This consistency check is especially important for
$\theta_\E$ measurements derived from microlens astrometry. The
\WFIRST\, microlensing mission will be the first time the astrometric
microlensing effect will be measured for large numbers of
events. Hence, the systematics inherent in this method are completely
unknown. The independent prediction of the lens mass from its measured
flux and parallax will provide a means to test the astrometric
measurements and vet for any systematics.

\section{Summary}

I have shown how microlens parallax measurements may be combined with
a measurement of the lens flux to yield a measurement of the lens mass
and distance. This method is particularly important for measuring
masses of lenses without measurements of $\theta_\E$, a category which
includes almost all single lenses. It is also relevant for low-mass,
nearby lenses for which the mass-distance relations derived from flux
and from $\theta_\E$ are partially degenerate.

Although in the past, many events of interest were more likely to have
$\theta_\E$ measurements as opposed to $\pi_\E$ measurements, that
situation is changing. First, that bias was partly a selection effect
in the publication record. Second, the \Spitzer\, and \K2\,
microlensing campaigns are enabling highly precise parallax
measurements for microlensing events. In addition, higher cadence,
higher precision microlensing surveys allow the detection of more
subtle planetary signals, so measurements of $\theta_\E$ are no longer
guaranteed for those events \citep{Zhu14}. For these systems,
combining parallax and flux measurements is the only way to measure
the lens mass. Finally, these issues will become even more acute for
the \WFIRST\, microlensing mission, which will routinely measure
parallaxes and lens fluxes for a significant fraction of microlensing
events.

\acknowledgements The author would like to thank Andy Gould for
helpful comments and discussion. Work by J.C.Y. was performed under
contract with the California Institute of Technology (Caltech) / Jet
Propulsion Laboratory (JPL) funded by NASA through the Sagan
Fellowship Program executed by the NASA Exoplanet Science Institute.


\end{document}